\documentclass[preprint,authoryear,round]{imsart}

\RequirePackage[OT1]{fontenc}
\RequirePackage{amsthm,amsmath}
\RequirePackage[numbers]{natbib}
\RequirePackage[colorlinks,citecolor=blue,urlcolor=blue]{hyperref}
\usepackage{amssymb}
\usepackage{graphicx}
\usepackage{url}
\usepackage{epsfig}
\usepackage{algorithm}
\usepackage{algorithmic}
\usepackage{amsmath,amssymb,amsthm}
\usepackage{natbib}
\usepackage{hyperref}
\usepackage{color}
\usepackage{subcaption}
\usepackage{verbatim}
\usepackage{marginnote}
\usepackage{bibentry}
\usepackage[normalem]{ulem}
\usepackage{xifthen}

\usepackage{sectsty}
\allsectionsfont{\sffamily\mdseries\upshape}

\usepackage[nottoc,notlof,notlot]{tocbibind} 
\usepackage[titles,subfigure]{tocloft}

\graphicspath {{Code/}}

\begin{document}
	
		\begin{frontmatter}
			\title{Coauthorship and citation networks for statisticians: Comment
				}
			\runtitle{Coauthorship and Citation Networks}
			
			\author{Vishesh Karwa \and  Sonja Petrovi\'c}
			\footnote{V. Karwa is with Department of Statistics, Harvard University; S. Petrovi\'c with Department of Applied Mathematics, Illinois Institute of Technology. 
			Preparation of this comment was supported in part by the U.S. Air Force Office of Scientific Research  Grant \#FA9550-14-1-0141 to Illinois Institute of Technology and by the Singapore National Research Foundation under its International Research Centre Singapore Funding Initiative and administered by the IDM Programme Office through a grant
			for the joint Carnegie Mellon/Singapore Management University Living Analytics Research Centre. 
			The authors are grateful to %their  collaborators, in particular to 
			Stephen Fienberg 		for endless motivation and support.}
			\runauthor{V. Karwa \and  S. Petrovi\'c}
	\end{frontmatter}	
\maketitle
%\tableofcontents

%%%%%============%%%%%%%%%%============%%%%%
\begin{comment}
\textcolor{cyan}{
 RECALL THE EMAIL FROM EDO: 
\\We are aiming to receive discussions of the following lengths, including references, in IMS AOAS format:
\\- up to 3 pages, for technical commentary
\\- up to 6 pages, for commentary that includes reanalyses; figures will not be counted toward the total length of the discussion.
}
\\
\end{comment}
%%%%%============%%%%%%%%%%============%%%%%

\section{Introduction}

%%%%%============%%%%%%%%%%============%%%%%
\begin{comment}
 (i.e., Summary of Ji and Jin's observations)}
{\color{red} I am trying to copy Steve's JASA comments to start the intro: Say something nice about the paper and get into the discussion :P If there a better way please let me know!}
\\
 yes, a good general strategy. 
\\ 
 SONJA IN FLIGHT: 
 I tried to get straight into the `flavor' of what we do and why right away. hope this works!!!
\end{comment}
%%%%%============%%%%%%%%%%============%%%%%

Analyses of coauthorship and citation networks  offer a fertile ground for studying research and collaboration patterns of  scientific communities. %
% THEY ALREADY SAY THIS EXACT THING:
% Such data have been collected and studied for several  fields, but not statistics before [[[{\color{cyan}edit sentence cut}]]]
%
%These types of networks have been collected and studied for several different fields, but not statistics. 
%but as Ji and Jin point out, such a dataset for statisticians did not exist before their work and that it is us who would be most interested in collecting and analyzing such data. 
Ji and Jin's efforts of collecting, cleaning and summarizing in various ways  citation and coauthorship networks for statisticians is a great step forward to provide the  community with % not only with
  a first such data set for self-study. 
%
%  , but also one whose connectivity structure inspires the development of new statistical models. 
%
% and model testing frameworks [....is this clear or confusing??]. 
%to close this gap. 
%\textcolor{cyan}{eh, sonja, sounds a little discombobulated  at the moment - try repackaging again! Maybe just remove the "but also one" part of the previous sentence -it is all said below anyway!}
They  perform several descriptive analyses of the underlying networks %underlying the coauthorship and citation data 
 to extract interesting patterns: they study  trends of productivity over time, extract  most prolific authors and research areas using various centrality measures, and  find communities in these networks.  
We look forward to seeing this dataset serving 
% \textcolor{blue}{OK hello, this is now repetitive -- see 2nd sentence!} 
% not only as a fertile ground for studying coauthorship and citation patterns in the field,  but also 
as a yardstick  for fitting social network models to large datasets. 
Perhaps more interestingly, we see it as raising  new research questions  from the modeling, data representation and computational points of view and  becoming a standard testbed for evaluating network models -- both old and new -- and testing scalability of inference procedures. 
 In this regard, it is with great  pleasure that we write this comment.

Here we take a model-based approach and consider the effects of various types of author interactions on the analysis and  inference about the citation and coauthorship datasets. We are generally  interested in three types of questions, two of which we discuss here: what are well-fitting models for the data? Is a simple network representation is best for answering questions we ask, or should we be considering alternative representations?  How can we scale existing network model fitting and goodness-of-fit testing procedures to networks of this size, as well as larger networks that the authors intend to collect? These  forthcoming data sets should reduce sampling bias, but of course come at a price of a dramatic increase in network size and computational cost.  We expect  that availability of the datasets Ji and Jin have provided the community  will encourage methodological research to push the limits of performing non-asymptotic inference in large and sparse networks.

We became aware of their data collection effort at a time when we were developing a basic exponential family model for hypergraphs, placing probabilities on occurrence of connected groups of nodes of arbitrary size  instead of pairs of nodes in a random graph. Indeed, in \cite{betaHypergraphs}, a mock example of a coauthorship dataset is used as a motivation for the new model. 
% whose sufficient statistics are still node degrees but they are calculated using hypergraphs. [[TOO MUCH INFO FOR INTRO.]]
Subsequently,  we introduced an ERGM that is based on a summary of a global connectivity structure called $k$-core decomposition \citep{shellERGM}. 
%%  [[[ \textcolor{cyan}{missing last sentence !!!  HOW ABOUT: }]]]  
Hence, we  reconsider Ji and Jin's data through the lens of these two network models, but first let us begin with  some more classical  models.

\section{Fitting  
dyad-independent models based on   node degrees}

%%%%%============%%%%%%%%%%============%%%%%
\begin{comment}
{\color{red} Results from fitting $p_1$ model to the citation network of authors. In the code, this is the citAdj}
\textcolor{blue}{for results, don't forget to re-read and incorporate section 5 of the "notes" tex file (in particular items 1,2,and 6), where we also already drafted some conclusions after a long discussion.} 
\textcolor{red}{I think you mean items 1,2, and 5? I already added notes from 1. 2 is no longer valid, since the $p_1$ model with differential reciprocity did not fit. and 6 talks about how the coauthor network is not doing what we think it does, I added this now.}
\\ \textcolor{blue}{add here briefly:} if The collected data is presented in forms of a simple graph or network: .....[describe briefly what nodes and edges mean in each network we are studying]. If there are multiple edges, thresholding is used....If several people wrote coauthored a paper, an edge between them exists in the coauthorship network. 
 \\ \textcolor{red}{I made an attempt...}
\\ GREAT. I worked off of this attempt - thanks for starting it! :) 
\end{comment}
%%%%%============%%%%%%%%%%============%%%%%

The data collected by Ji and Jin contains  two key datasets: (1) a bipartite graph of authors and papers where a link exists from node $i$ to node $j$ if author $i$ wrote paper $j$; (2) a network of citations of papers where a link exists from node $i$ to node $j$ if paper $i$ cites paper $j$. 
From these two datasets, Ji and Jin extract two networks whose nodes are authors: two coauthorship networks  and one citation network. 
In the ``Coauthorship network A'' there is an undirected edge between nodes $i$ and $j$ if author $i$ coauthored at least $2$ papers with author $j$.
In the ``Coauthorship network B'' there is an undirected edge between nodes $i$ and $j$ if author $i$ coauthored at least $1$ paper with author $j$. In the citation network of authors, there is a directed edge from author $i$ to author $j$ if  $i$ has cited  at least $1$ paper by  $j$.
	
It is important to note that, strictly speaking, the number of citations between authors and the number of coauthors are counts greater than $1$. They are converted to a binary network by using thresholding, a popular technique in network analysis used to avoid multiple edges. Moreover, in the citation network of authors, the self citations are set to $0$ to avoid  loops in the network representation, so that the result is  a desirable simple graph. 
%%%%%============%%%%%%%%%%============%%%%%
\begin{comment}
\textcolor{cyan}{[[[ Qn: Are we still happy with having these 2 paragraphs summarizing data structure inside this section? just checking if we like the flow of text now. I am thinking it is fine.]]]} 
\textcolor{red}{[[I think this should be enough!]]}
%% There are several ways to represent the data - the two common structures being a network (bipartite, directed, or undirected), and a contingency table. Depending on the representation, a different analysis can be carried out - that is the type of model that can be fitted to the data depends on the representation.

\textcolor{red}{I feel like the flow is missing from the previous para to this one. See if you like my modification}
\textcolor{blue}{I agree - this is great}
\medskip
\end{comment}
%%%%%============%%%%%%%%%%============%%%%%

Once a network representation is extracted from the data, there are many ways to analyze it using descriptive statistics. 
 For instance,  Ji and Jin consider degree centrality of these extracted networks to measure the importance of nodes. 
In  the coauthorship network, they use node degrees to identify most collaborative authors, while in the citation network of authors, they use the number of citers -- the in-degree of the corresponding directed graph -- to identify top authors. 
 From the point of view of modeling, 
  it is natural to ask whether  degree-based analysis is sufficient for these networks; in particular, models based on degrees exhibit dyadic independence and we question whether such an assumption is valid. 
In other words, when a statistic is used to summarize a dataset, 
 we see it as a sufficient statistic of some model and then ask what that model is and how well it fits the data. 
%%%%%============%%%%%%%%%%============%%%%%
\begin{comment}
\textcolor{cyan}{[[[i hope this last sentence is not redundant and is ok, or do we need to remove it? but if we remove it, we should rephrase the next sentence, as it currently refers to "this question" for a specific summary statistic:]]]} 
\textcolor{red}{ I think the last sentence os ok, but perhaps we should remove the phrase subsequently used for sampling? As Jin and Ji dont do any sampling. I think we should say that Jin and Ji do a descriptive analysis of the networks using degrees. This naturally leads to the question that are degrees good summary statistics of the network? A simple way to answer this is to consider models where degree sequence is a sufficient statistic and check if such a model provides an adequate description of the data. If so, degree based descriptive analysis makes sense. This is the approach we take..}
\textcolor{blue}{ i understand, you are right --- perfect! } 
%  
% To this end, we perform goodness-of-fit tests for two degree-based models: the $\beta$ model for random undirected graphs \cite{}, and the $p_1$ model for random directed graphs \citep{HL81,FW81}. 
%
\end{comment}
%%%%%============%%%%%%%%%%============%%%%%

We investigate the above question for node degrees  of both citation and coauthorship networks. 
%, since   Ji and Jin use node degrees to identify most prolific and most cited authors. 
The simplest but nicely interpretative model for random directed graphs whose sufficient statistics are the node in- and out-degrees  is the $p_1$ model \citep{HL81}. 
The model assumes that dyads (i.e., citation pairs) are independent and assigns  probabilities to  four types of citations: outgoing, incoming, reciprocated, and none,  represented  by $i \rightarrow j$, $i \leftarrow j$,  $i \leftrightarrow j$, and $ i \not\sim j$, respectively. Node-specific parameters represent attractiveness and expansiveness, and there is an additional parameter for the overall tendency of the network to reciprocate citations. This parameter can be set to be zero or a nonzero constant; Holland and Leinhardt consider both versions of the model. 
\cite{FW81} represent $p_1$ in log-linear form, turning networks into $0/1$ contingency tables, and extend the model to allow for the differential reciprocity effect by including  dyad-specific reciprocation parameters. (The model was later extended to fit within the block model framework; see \cite{fienberg1985statistical}.) 
%% {\color{blue}GOTTA LOCATE THAT PAPER!!... Are you thinking about this one? -- YES  thanks!! had no internet in flight :) } 
For undirected graphs such as the coauthorship network, $p_1$ reduces to the $\beta$ model \citep{BlitzDiac10} that has had a long history in various literatures. 
%  literature, see for example recent work  \cite{CDS11} %, \cite{RPF:11},  % \cite{Psurvey} 
% and references give therein. 
%%%  \textcolor{red}{$\leftarrow$ decide 1-2 best references to put here. The model is much older than Chat-Diac-Sly! Add vishesh+sesa AOS paper? add/remove \cite{RPF:11}?? not sure. why not - AOS is a good thing to cite hehe}

The log-linear representation allows the use of tools from algebraic statistics introduced in \cite{DS98} to fit the $p_1$ model and perform 
a non-asymptotic goodness-of-fit test; see  \cite{PRF10} for the basic theoretical  results  for the $p_1$ model  and  \cite{GPS14} for generalizations and implementation. 
  %%  {will add just a couple of sentences saying how this is done, that it is an exact test using mcmc to sample from the fiber.} -- no need -- it is below. 

\paragraph{Results of goodness-of-fit tests.} 

We perform an exact test of model fit for the $p_1$ model with dyad-dependent reciprocation (the most general version)  to the largest connected component of the citation network of authors. 
%% \textcolor{cyan}{[[by the way J\&J do the same thing - take the `giant component' - in their paper. Just for the record! ]]}
The test is done by running the Markov chain  from \cite{GPS14}. After  $n=100000$ steps, the resulting $p$-value is $0.007194245$. This result indicates that the $p_1$ model does not fit the citation network of authors. 

The lack of fit of the $p_1$ model suggests  that the network  of  citations  may have tendencies to be transitive and the dyads may  not be independent.
%%%%%============%%%%%%%%%%============%%%%%
\begin{comment}
[[[ 
	 .... we are actually hinting at the fact that maybe, just maybe, using degrees to identify most cited authors is not so great. but i don't want to say that explicitly. because it sounds weird. I mean, obviously if you want to see how is MOST cited, you just count. But we should watch for only accidentally stumbling into concluding that  although counting citations is a natural way to rank people, it seems that there is no statistical evidence that that's a good summary of features of this network!! ... um, choose words carefully; hehe - the issue is probably going to raise a bunch of discussions (specially if you mention it at JSM)!... 
]]]\\
 Hence the bi-directed degree sequences 
[TECHNICAL WORDING: bi-degree sequence OR directed degree sequence OR in- and out-degrees] 
are not good summary statistics and we need to look for other measures of centrality, that also capture triadic effects.
\end{comment}
%%%%%============%%%%%%%%%%============%%%%%
While counting citations is a natural way to perform rankings, such a count -- corresponding to the directed degree sequence of the citation network -- does not offer a good summary statistic for the citation network data. Instead, we need to look for other more complex measures of centrality, for example those that are also able to capture triadic  or transitive effects. 

 %If the model fits, the results will indicate that the dyadic independent model is a reasonable fit, and that degree sequences are a good summary statistic. The network of citations and coauthorship are not cliquish. This can be interpreted as follows: In the citation network of papers, if paper $i$ cites $j$ and paper $j$ cites $k$, it is not necessary that $k$ will cite $i$. That is there is no triadic closure. In the author citation network, this will mean that if author $i$ cites author $j$ and if author $j$ cites author $k$, author $k$ need not cite author $i$. On the other hand, if the $p_1$ model does not fit, it means that the degree sequences are not good summary statistics and we need to look at something more complicated. (such as the cores or the triangles?) and that the network of citations and author citations may be cliquish!  This also means that looking at the degree to identify node importance in the networks is not enough (e.g. table 2 is used to identify important people in various networks by looking at the corresponding degree. But if the network is cliquish, degrees are not enough.)

%\paragraph{Reciprocity of Citations.} 
Another comment is in order. The variants of the  $p_1$ model are naturally set up to test the reciprocation effect: 
%There is a directed edge from node $i$ to node $j$ if author $i$ cites author $j$ at least once. We consider the following question: Do 
do authors reciprocate citations? That is, if author $i$ cites author $j$, is author $j$ likely to cite a paper by author $i$? The lack of fit of the model that captures the reciprocation effect means that the answer to this question is `no', however, it does not directly imply that there are no transitive effects of other kinds that we are unable to test at the moment. 
%%%%%============%%%%%%%%%%============%%%%%
\begin{comment}
This question can be answered by fitting a $p_1$ model  with edge dependent reciprocation to the citation network of authors. \textcolor{blue}{ I don't know what happened here?} 
\textcolor{red}{Since th $p_1$ with edge dependent reciprocation and the $p_1$ with $0$ reciprocation both did not fit, can we conclude that there is no reciprocity? Or we can only conclude that the $p_1$ model did not fit. The point is that there may be reciprocity in the data, but we may need a transitive model to detect it. A model that is dyad dependent. I was trying to fit an ergm with triangles and mutuality, but it did not converge. This is why I did not write anything here :P Perhaps we can remove this point altogether?}
\textcolor{blue}{aha, i see! OK. No, let's KEEP  the point and summarize in 2 sentences that the reciprocity effect is rejected [we can say that for sure for now] but there may be OTHER KINDS of transitive effects [different from just 'i reciprocate citation!' which anyway we were sort of hoping wasn't true right?hehe -- and we leave it at that as an open problem for people to work out and cite us.} 
\\\textcolor{cyan}{
Will attempt to summarize this on Fri morning.
}
\end{comment}
%%%%%============%%%%%%%%%%============%%%%%

\smallskip

Similarly, we perform an exact test of model  fit of the $\beta$ model to the largest connected component of the coauthorship network $A$. The $p$-value from the goodness-of-fit test obtained by running   the Markov chain  on $n=100000$ steps is $0.997$, indicating  a pretty good fit. The dyads in the coauthorship network can be assumed to be independent, and the network does not have any triadic closure effects; i.e., if author $i$ wrote a paper with author $j$, and author $j$ wrote a paper with author $k$, then it is not necessary that author $i$ has written a paper with author $k$. This result is somewhat surprising since experience suggests that collaborators of an author $i$ may tend to collaborate with each other, including $i$. On a closer look, perhaps it is less surprising:  forgetting the structure of the original data set and converting it to the underlying graph (by recording only the dyadic relationships) results in independence and node degrees being a good summary of the graph. 
%\textcolor{cyan}{adding - but need to rephrase [not to be colloquial]:}
%On the other hand, there is something nice going on: $\beta$ model fitting well means the degrees of nodes are a good summary of the data (so, we need not focus on dyad independence only) -- but of course this statement holds not for the original data, but rather for the thresholded underlying graph supporting the original data. 

\section{What lies beyond node degrees?} 
 %degree-centric analyses [i don't like "centric"] \textcolor{red}{Ok we can use something else!}Unable to come up with the section title right now.}
As we saw above, the $p_1$ model based on node degrees does not fit the citation network and, hence, degree-based analyses may be of suspect. It is well-known that, in general, degree-based models may fail to capture certain vital connectivity information about the network.  
In applications such as the present one, we  may be interested in  the type of \emph{global} connectedness  effectively captured by the
 {cores decomposition} of a graph introduced by \cite{Seidman83} (see \cite{shellERGM}   for statistical considerations).
 For the directed citation network, we compute the $k$-core using the in-degree which measures the number of times an author is cited. Intuitively, the $k$-core captures the innermost core of ``highly cited'' authors. To be in the innermost core, it is not sufficient to have the highest number of citations, but one must receive citations from authors who are themselves cited by many.

To convert the citation counts between authors to a directed network of author citations, we use varying threshold values $c$. A directed edge exists from node $i$ to node $j$ if author $i$ cites author $j$ at least $c$ times. We consider $c = \{1,2,3,4,5\}$, providing  $5$ different networks. Table~\ref{tab:coretopk} shows results of selecting the top $5$ highly cited authors based on their degree in the respective cores. 

\begin{table}[ht]
	\centering
	\caption{Top $5$ authors selected using the $k$-core decomposition with different values of $c$}
	\begin{tabular}{lllll}
		\hline
		1 & 2 & 3 & 4 & 5 \\ 
		\hline
		Jianqing Fan & Jianqing Fan & Jianqing Fan & Jianqing Fan & Peter Hall \\ 
		Hui Zou & Hui Zou & Hui Zou & Hui Zou & Hans-Georg Muller \\ 
		Ming Yuan & Peter Buhlmann & Peter Hall & Runze Li & Raymond J Carroll \\ 
		Peter Buhlmann & Cun-Hui Zhang & Runze Li & Peter Hall & Fang Yao \\ 
		Runze Li & Runze Li & Raymond J Carroll & Hans-Georg Muller & Jianqing Fan \\ 
		\hline
	\end{tabular}
\label{tab:coretopk}
\end{table}

Compare the results of Table \ref{tab:coretopk} to Column 3 of Table 2, where Ji and Jin identified the top 3 most cited authors,``Jianqing Fan'', ``Hui Zou'', ``Peter Hall'', using the in-degree. When $c \in \{1,2,3,4\}$ ``Jianqing Fan'' and ``Hui Zou'' appear as the top 2 authors, and Peter Hall appears in the third place only when $c=3$. However, when $c =5$, Peter Hall is the most cited author. Our goal here is to simply illustrate the point the results depend on the type of centrality measure chosen, and the threshold used to create the network.

\begin{figure}[t]
		\begin{center}
			\includegraphics[width=0.6\textwidth]{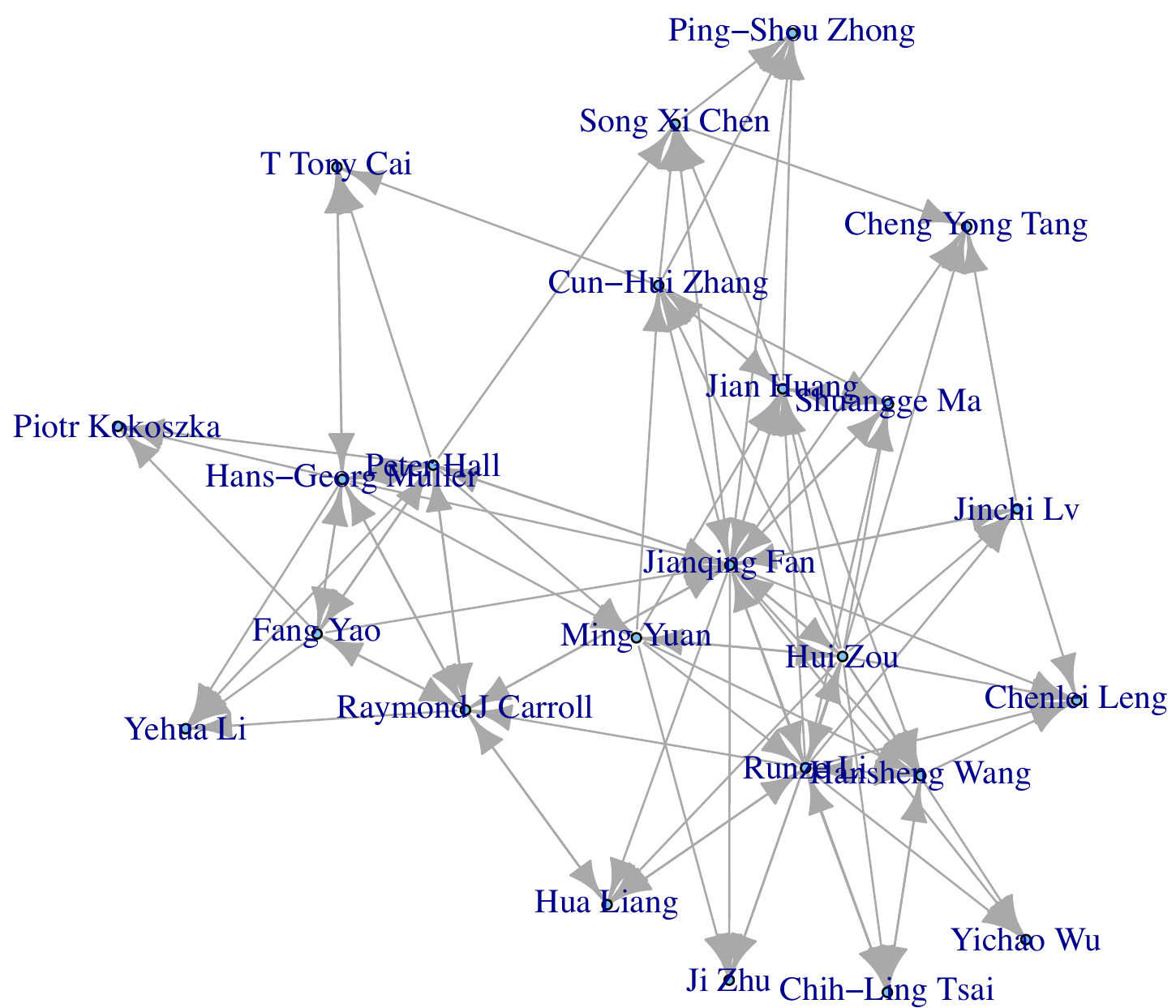}
			\caption{The innermost core of the citation network where an edge exists if there are at least $4$ citations.
			  %\textcolor{red}{TO DO: Great,  but this looks undirected! --  make the arrowheads bigger so we can see them!}
			  }
			\label{fig:innerCoreCitation}
		\end{center}
	\end{figure}

As an illustration, Figure \ref{fig:innerCoreCitation} shows the innermost core of the citation network of authors when $c=4$.

\begin{figure}[t]
	\begin{center}
		\includegraphics[width=0.4\textwidth]{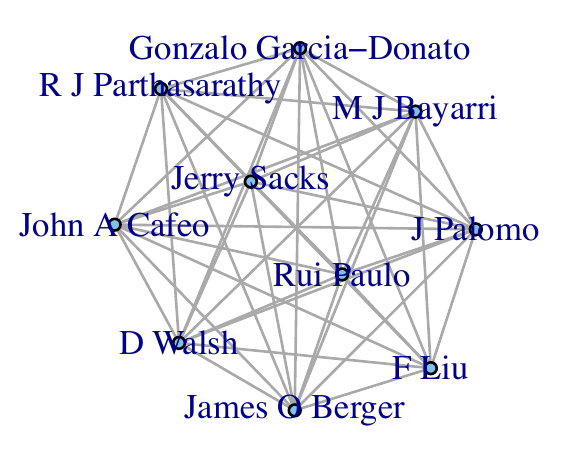}
		\includegraphics[width=0.47\textwidth]{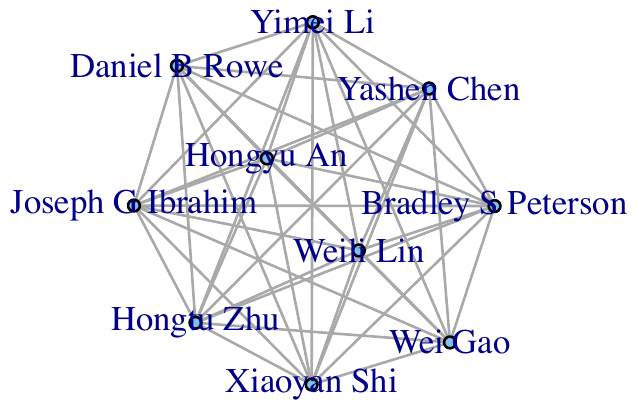}
		\caption{The innermost core of coauthorship network A.
		%\textcolor{red}{CITATION? COAUTHOR? fix caption.}  Citation Network of authors. An edge exists if there are at least $3$ citations. -- no, at least 1 joint paper. \textcolor{red}{ [note again: graphics don't count toward page limit!]}
		}
		\label{fig:innerCoreAuthor}
	\end{center}
\end{figure}

\smallskip
The $k$-core decomposition of the coauthorship network is also instructive. For this case, we perform a core decomposition of the coauthorship network $A$.  Figure \ref{fig:innerCoreAuthor} shows the inner most core of the coauthorship network. The innermost core of the coauthorship network consists of two connected components and every node has degree 9! In other words, the innermost core consists of 2 cliques of size $10$. On further exploration, it turns out that the two cliques correspond to two papers, \cite{bayarri2007computer} and \cite{zhu2009regression}, each with $10$ authors. 
This example   illustrates that a network representation of the coauthorship data can be misleading: it is not possible to distinguish between the cases of authors writing many joint papers and when many authors  writing  one. 

%\bibentry{bayarri2007computer}.

\section{Need for new models and representations} 

Both of the previous two sections %, Results from the previous two sections (last sentence of sec.2 and also of sec.3!) 
motivate recording  higher-order interactions from the data. 
In addition, Table~\ref{tab:coretopk} %[motivating example] 
suggests the thresholding is not good as it looses information. 
%%%%%============%%%%%%%%%%============%%%%%
\begin{comment}
{\color{red} (1) We can say here that graphs are not the best ways to represent this data. We can use the previous paragrah to motivate why hypergraphs make sense. (2) We can say that thresholding is not good, as it looses information, again use table 1 as a motivating example. We need to think in terms of counts, and contingency tables. Finally (3) We need new models for both these representations, for hypergraphs and for contingency tables.} -- GREAT. 
I can also add some general stuff here --Sonja.\\
{\color{red} Vishesh's attempt to add notes on data representations:}
\end{comment}
%%%%%============%%%%%%%%%%============%%%%%
There are several ways to represent the data, two  common structures being a network (undirected, directed or bipartite)  and a contingency table, each allowing for different analyses to be carried out. That is, the type of model that can be fitted to the data depends on the representation. 

\paragraph{Contingency table  representation.} For  $I$ authors, $J$ research areas and $K$ journals, consider an $I \times I \times J \times K$ contingency table where the $(i, i', j, k)$ entry counts the number of times author $i$ cites author $i'$ in research area $j$ and journal $k$. A similar representation can be obtained for the coauthorship network, where  we count the number of times author $i$ and author $j$ wrote a joint paper. These representations preserve the citation and coauthorship count data.  %allow the count data to remain in the count form. 
We can then collapse the table to an $I \times I$ author-by-author table and fit log linear models to the citation counts. 
In essence, we seek to avoid thresholding, as in  the generalized $\beta$ model  discussed by \cite{RPF:11} for weighted networks represented in table form. 
%%%    This will be a more general representation of the coauthorship network. Ji and Jin convert this contingency table to a Citation network by thresholding: All counts greater than $1$ are considered to be a directed edge \textcolor{cyan}{i think you can do the exact same thing for the coauthorship data and the beta model} . We believe that there is no need to threshold the counts and the data can be analyzed directly as a contingency table. 

\begin{figure}[t]
	\begin{center}
		\includegraphics[width=0.55\textwidth]{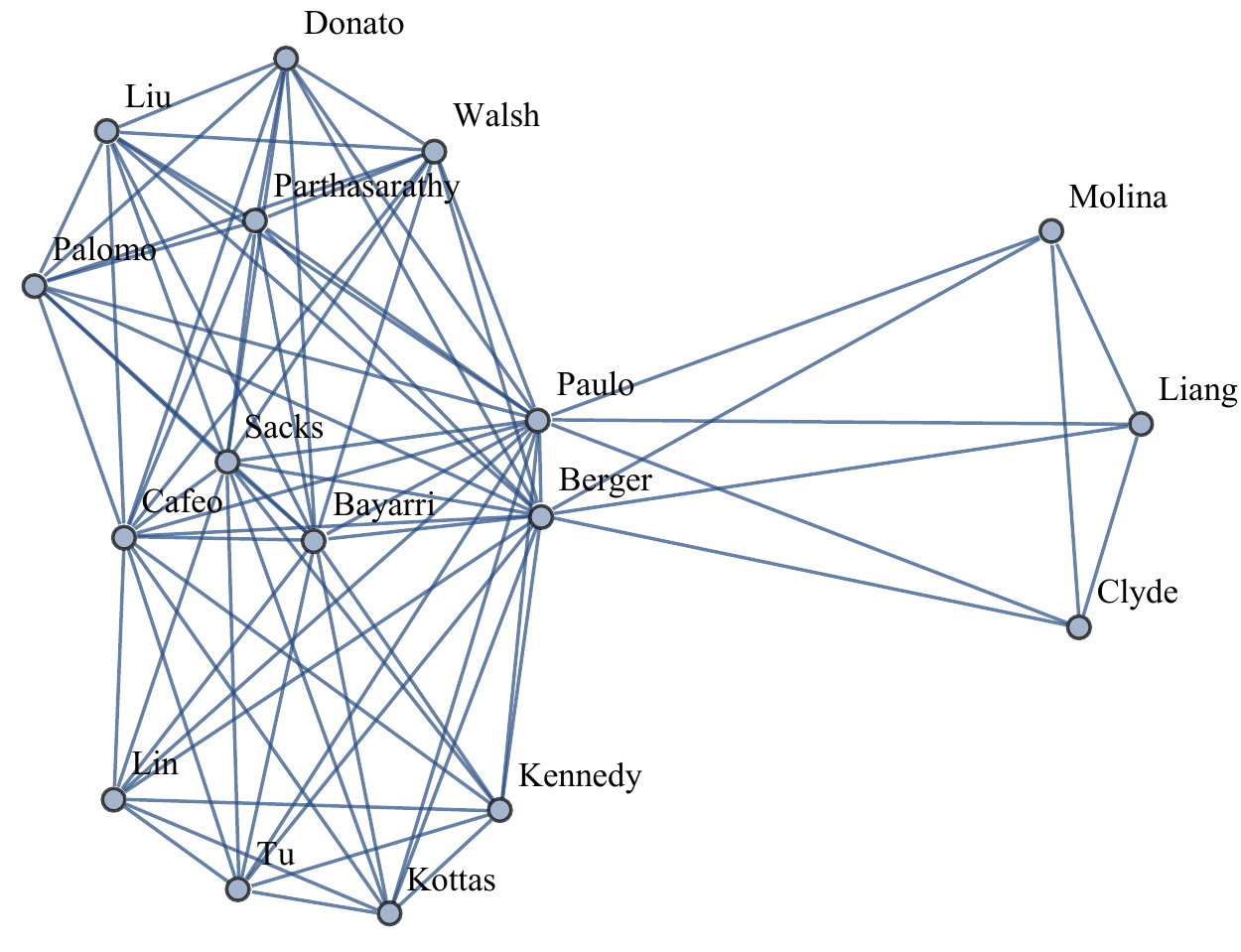}
		\includegraphics[width=0.4\textwidth]{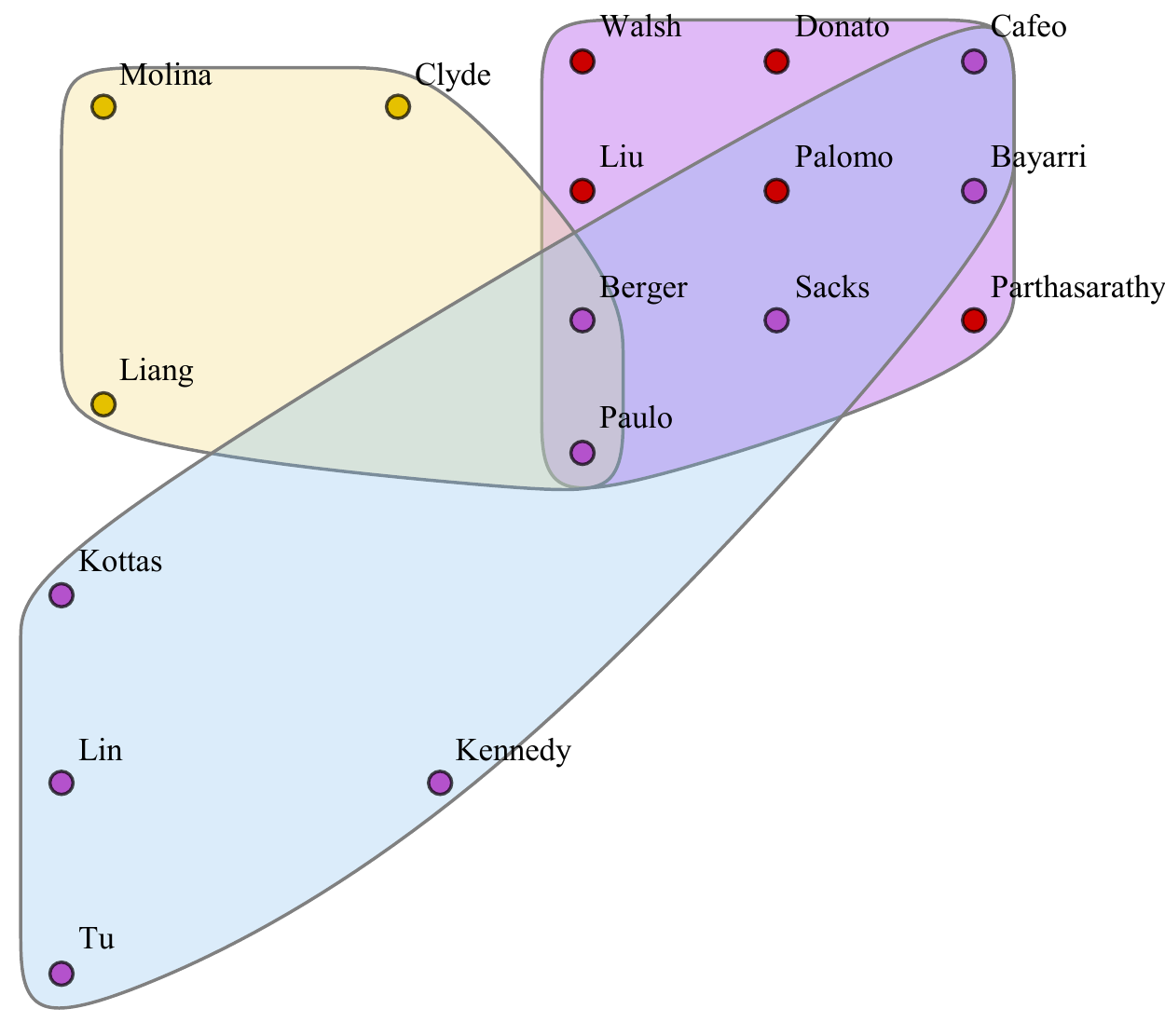}
		\caption{The graph and hypergraph representation of a subnetwork of the co-authorship network A. The comparison clearly shows the loss of information in representing a hyperegde by edges in the network.
			%\textcolor{red}{CITATION? COAUTHOR? fix caption.}  Citation Network of authors. An edge exists if there are at least $3$ citations. -- no, at least 1 joint paper. \textcolor{red}{ [note again: graphics don't count toward page limit!]}
		}
		\label{fig:hypergraphvsNetwork}
	\end{center}
\end{figure}

\begin{figure}[t]
	\begin{center}
		\includegraphics[width=0.9\textwidth]{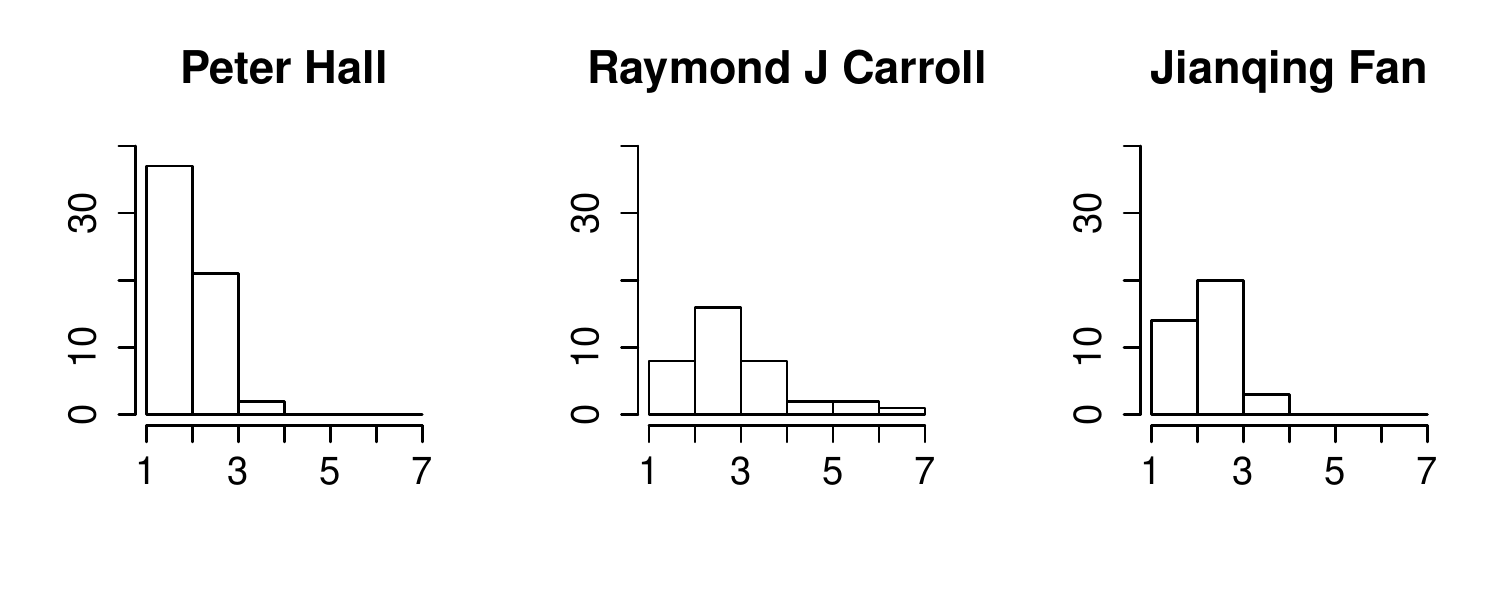}
		\caption{Histograms of hypergraph degrees with respect to edge size in the coauthorship Hypergraph. The top $3$ authors by hypergraph degree are shown.
			%\textcolor{red}{CITATION? COAUTHOR? fix caption.}  Citation Network of authors. An edge exists if there are at least $3$ citations. -- no, at least 1 joint paper. \textcolor{red}{ [note again: graphics don't count toward page limit!]}
		}
		\label{fig:hyperDegreeHist}
	\end{center}
\end{figure}

\paragraph{Hypergraph representation.} 
Coauthorship networks may not be measuring what they intend to measure; recall illustrative examples above. 
%%%%%============%%%%%%%%%%============%%%%%
\begin{comment}
 For example, we looked at the innermost core of the coauthorship network and found two disjoint cliques of size $10$. These cliques correspond to authors that wrote a single paper together. Thus it is possible to be in the innermost core of the coauthorship network, just by writing a single paper with many authors. This may be ok, but if one is interested in measuring the number of papers written by an author and how collaborative an author is, this is not a good measure. For instance, if one writes a single paper with 100 authors, she would be deemed most collaborative in the coauthorship network. There is no way to distinguish between the case when an author writes many papers with few authors vs an author who writes a single paper with many authors and so on. They all get grouped as high degree nodes in the coauthorship network. We believe that these problems can be resolved by using a hypergraph representation.
\end{comment}
%%%%%============%%%%%%%%%%============%%%%%
To prevent information loss and model higher-order interactions, we represent the raw coauthorship data via a hypergraph, which is a generalization of a graph. A random hypergraph is a collection of random hyperedges which are occurrence of groups of nodes of arbitrary size $k$.  For example, a hyperedge (for simplicity also called an edge) of size $k$ containing nodes $i_1,\dots,i_k$ exists if authors $i_1,\dots,i_k$ wrote a joint paper.  
Figures~\ref{fig:hypergraphvsNetwork}, \ref{fig:hyperDegreeHist}, and Table~\ref{tab:hyperdegreetopk} highlight different aspects of the data that can be extracted from the  hypergraph representation. 

\cite{betaHypergraphs} introduce  $\beta$ models for random hypergraphs. \cite{simonhypergraph} give a geometric representation of hypergraphs. Clearly, more complex statistical models for  random hypergraphs %-- random occurrence of groups of nodes of arbitrary size -- 
are necessary, as the degree-based $\beta$ model is sure to have similar shortcomings on hypergraphs as it did on graphs. 
In addition, Figure \ref{fig:hyperDegreeHist}  suggests placing heterogeneous weights on hypergraph degrees with respect to edge size. 
Furthermore, we may wish to preserve edge multiplicities representing multiple joint papers by same groups in contingency table form here as well. 

\begin{table}[ht]
	\centering
	\caption{Top $3$ authors that have $k \in \{1,2,3,4,5\}$ or more collaborators (including themselves), based on the hypergraph representation}
	%\scriptsize
	\begin{tabular}{lllll}
		\hline
		1 & 2 & 3 & 4 & 5 \\ 
		\hline
		Peter Hall 			& Peter Hall 		& Raymond J Carroll & Raymond J Carroll & Joseph G Ibrahim \\ 
		Raymond J Carroll 	& Raymond J Carroll & Peter Hall 		& Joseph G Ibrahim & Raymond J Carroll \\ 
		Jianqing Fan 		& Jianqing Fan  	& Jianqing Fan 		& Hongtu Zhu & Hongtu Zhu \\ 
		\hline
	\end{tabular}
	\label{tab:hyperdegreetopk}
\end{table}

%%%%%============%%%%%%%%%%============%%%%%
		\begin{comment}
{\color{red} Add more here! E.g a picture highlighting this difference, also the point that we need more models for hypergraphs. Cite Sonja and Despina's paper etc.}
\textcolor{cyan}{
well, we are out of space, but once we include the pictures we talked about - from the real data - that will be illustrative enough! $:)$ 
}
\textcolor{blue}{IT WOULD BE JUST SO GREAT IF WE COULD AT LEAST PROVIDE HYPERGRAPH DEGREES AND RE-REPORT SOME OF THE OBSERVATIONS FROM THAT POINT OF VIEW - because with the beta hypergraph model we can't do the  GoF test yet, only the MLE right now. Hmm. I am pretty sure we can ask for degree of a node in a hypergraph in R!! just have to figure out the 2 lines of code we need. SHALL WE TRY?}
		\end{comment}
%%%%%============%%%%%%%%%%============%%%%%

%%%% CLOSING SENTENCE: 
%Let us close this comment by concluding that we need new statistical models for both hypergraphs and contingency table representations of coauthorship and citation data. 

\bibliographystyle{named}
\bibliography{ref,AlgStatNtwks,kCoresERGM}

\end{document}